\documentclass[fleqn,usenatbib]{mnras}  

\usepackage[T1]{fontenc}
\usepackage{ae,aecompl}

\usepackage{mathptmx}

\usepackage{graphicx}	
\usepackage{amsmath}	
\usepackage{amssymb}	
\usepackage[british]{babel}
\usepackage[]{units}
\usepackage{etoolbox}
\usepackage{mnrasArxivFix}
\hypersetup{pdfauthor={Errani et al.}, pdftitle={The effect of a disc on the population of cuspy and cored dark matter substructures in Milky Way-like galaxies},
            pdfkeywords={galaxies: haloes, formation, structure, dwarf; cosmology: dark matter; methods: numerical}}
\title[Abundance of DM substructures in MW-like galaxies]{The effect of a disc on the population of cuspy and cored\\ dark matter substructures in Milky Way-like galaxies}

\author[Errani et al.]{Rapha\"el Errani$^{1}$\thanks{E-mail:
raer@roe.ac.uk}, Jorge Pe\~narrubia$^{1}$, Chervin F. P. Laporte$^{2}$ \& Facundo A. G\'omez$^{3}$
\\
{$^1$ Institute for Astronomy, University of Edinburgh, Royal Observatory, Blackford Hill, Edinburgh EH9 3HJ, UK}\\
{$^2$ Department of Astronomy, Columbia University, 550 W 120th St., New York, NY 10027, USA}\\
{$^3$ Max-Planck-Institut f\"ur Astrophysik, Karl-Schwarzschild-Str. 1, D-85748, Garching, Germany}
}

\date{\vspace*{-0.3cm}Accepted 2016 October 12. Received 2016 October 7; in original form 2016 August 5; Editorial Decision 2016 October 11}

\pubyear{2015}

\begin{document}
\label{firstpage}
\pagerange{\pageref{firstpage}--\pageref{lastpage}} \pubyear{2016}
\maketitle

\begin{abstract}
We use high-resolution $N$-body simulations to study the effect of a galactic disc on the dynamical evolution of dark matter substructures with orbits and structural parameters extracted from the Aquarius A-2 merger tree \citep{Springel2008}. Satellites are modelled as equilibrium $N$-body realizations of generalized Hernquist profiles with $2\times10^6$ particles and injected in the analytical evolving host potential at $z_\mathrm{infall}$, defined by the peak of their mass evolution.
We select all substructures with $M_{200}(z_\mathrm{infall})\geq\unit[10^8]{M_\odot}$ and first pericentric distances $r_p<r_{200}$.
Motivated by observations of Milky Way dwarf spheroidal galaxies, we also explore satellite models with cored dark matter profiles with a fixed core size $r_c=0.8\,a_s$ where $a_s$ is the Hernquist scale radius.
We find that models with cuspy satellites have twice as many surviving substructures at $z=0$ than their cored counterparts, and four times as many if we only consider those on orbits with $r_p\lesssim0.1\,r_{200}$. For a given profile, adding an evolving disc potential reduces the number of surviving substructures further by a factor of $\lesssim2$ for satellites on orbits which penetrate the disc ($r_p\lesssim\unit[20]{kpc}$). For large $r_p$, where tidal forces and the effect of the disc become negligible, the number of satellites per pericentre bin converges to similar values for all four models.
\end{abstract}



\section{Introduction}

Despite the success in modelling dark matter (DM) structure formation on large scales, the clustering properties of DM on galactic scales are fiercely debated. The  number of observed dwarf galaxies around the Milky Way (MW) is lower by many orders of magnitude than the number of \emph{cold} dark matter (CDM) substructures predicted by cosmological simulations - e.g. \citet{Moore1999} predict a number of CDM substructures with bound mass $\gtrsim\unit[10^8]{M_\odot}$ larger by a factor 50 than the number of observed MW satellites in this mass range, and \citet{DiemandMoore2005} predict a total of $10^{15}$ surviving CDM subhaloes down to the mass determined by the free-streaming length ($\sim\unit[10^{-6}]{M_\odot}$).
Within the CDM paradigm this can be explained by a suppression of galaxy formation in low-mass DM haloes, through the combined effect of inefficient cooling of gas during H\textsc{i} re-ionization by UV photo heating \citep{Gnedin2000} and supernova feedback \citep[e.g.][]{DekelSilk1986} leading to DM haloes that do not contain any stars. 
Alternatively DM particle properties, such as low-mass \emph{warm} DM \citep[e.g][]{Lovell2014}, can lead to a suppression of the small-scale power resulting in a reduction of substructures in galactic haloes.
Likewise the presence of a galactic disc enhances the disruption rate of substructures \citep[e.g.][]{DOnghia10, Penarrubia2010} and alters the amount of surviving substructures.
Several methods have been proposed to detect dark substructures through their gravitational interaction with baryons, e.g. surface brightness reconstruction of strong lensing arcs \citep{Vegetti2009}, the heating of tidal streams \citep[e.g.][]{Ibata2002heating}, or the disruption of wide binaries in dwarf spheroidal (dSph) galaxies \citep{PenarrubiaLudow2016} - these efforts still remain unsuccessful to date.

The inner structure of DM haloes is strongly linked with the number of DM substructures lingering in MW-like galaxies.
The efficiency of tidal stripping inside the host halo depends on the inner density distribution of the satellites. In particular, 
dwarf galaxies embedded within DM haloes with constant-density \emph{cores} are less resilient to tidal disruption than those embedded within centrally-divergent \emph{cuspy} haloes \citep{Penarrubia2010}.
While the density profiles of haloes in DM-only simulations are well fit by the cuspy \emph{NFW} profile \citep{nfw1997}, 
contradicting results based on the observations of MW satellites have provoked a critical debate on whether dwarf galaxies are cuspy (\citealt{Richardson2014}), or cored (\citealt{Walker2011}, \citealt{Amorisco2012}, \citealt{AmoriscoAgnelloEvans2013}), whereas dynamical modelling of local dwarf galaxies argues in favour of a range of cuspy and cored profiles \citep[][]{Oman2015}.
In addition, none of the observed bright MW satellites ($L_V>\unit[10^5]{L_\odot}$) has an estimated mean density within the stellar half-light radius consistent with the most massive cuspy subhaloes predicted by cosmological simulations \citep{BoylanKolchin2011}.
Different processes such as gravitational heating by baryonic clumps \citep[e.g.][]{Nipoti2014} and supernova feedback \citep[e.g.][]{Pontzen2012} have been proposed to remove the density cusp from the centres of dwarf galaxies.
However, recent studies taking into account the effects of tidal stripping on baryonic clumps show significantly milder changes to the central DM density slopes \citep{LaporteWhite2015}. Similarly, simple analytic energy arguments suggest that supernova feedback only affects the brightest dwarfs \citep{Penarrubia2012}, although stochastic star formation in low-mass haloes may help to ease the energetic constraints \citep{Read2016cores}.
Furthermore, within CDM, erased cusps may still regrow through minor mergers \citep{LaportePenarrubia2015}.
Alternative DM particle models which allow for a large enough cross-section for self-interaction \citep[e.g.][]{Spergel2000} or a small enough DM particle mass \citep[e.g.][]{TremaineGunn1979} could prevent cusps to form. Yet, \citet{Maccio2012} argue that a particle mass small enough to produce core sizes of $r_c\sim\unit[1]{kpc}$, as measured e.g. for the Fornax dwarf galaxy \citep{AmoriscoAgnelloEvans2013}, is in contradiction to the constraints set by the large scale structure and would prevent the dwarf galaxy to form in the first place. Other studies suggest that ultra-light axions may solve the missing satellites and cusp/core problems simultaneously \citep{Hu2000}.

In this Letter, we examine the tidal effect of an evolving galactic disc on the survival of DM substructures in MW-like haloes. We are particularly interested in the different dynamical evolution of satellites with respect to their internal DM density profile. 
Our contribution follows up the work of \citet{DOnghia10} who re-simulated the Aq-A-2 halo merger tree with an added analytical disc and found that the presence of a disc enhances the tidal disruption of substructures. More recent studies by \citet{YurinSpringel2015}, who inserted $N$-body discs in the Aq-A-5 to H-5 haloes, and \citet{ZhuMarinacci2016}, who compared DM-only to full-hydrodynamical simulations of the Aq-C-4 main halo, show similar results.
Motivated by the approach of \citet{Bullock2005}, we perform controlled simulations which follow the evolution of individual satellites extracted from the Aquarius Aq-A-2 merger tree \citep{Springel2008} in an analytical, evolving host halo, where we change single parameters at a time and quantify their effect on the surviving satellite population.
This Letter is structured as follows: Section \ref{sec:numMet} describes our re-simulation technique and the numerical models for the evolving host and satellite galaxies. We present the results of our controlled simulations in Section \ref{sec:ControlledSims}, and finally discuss our findings and plans for future work in Section \ref{sec:Conclusions}.


\begin{figure*}
  \centering 
  \begin{minipage}{\textwidth}
   \includegraphics{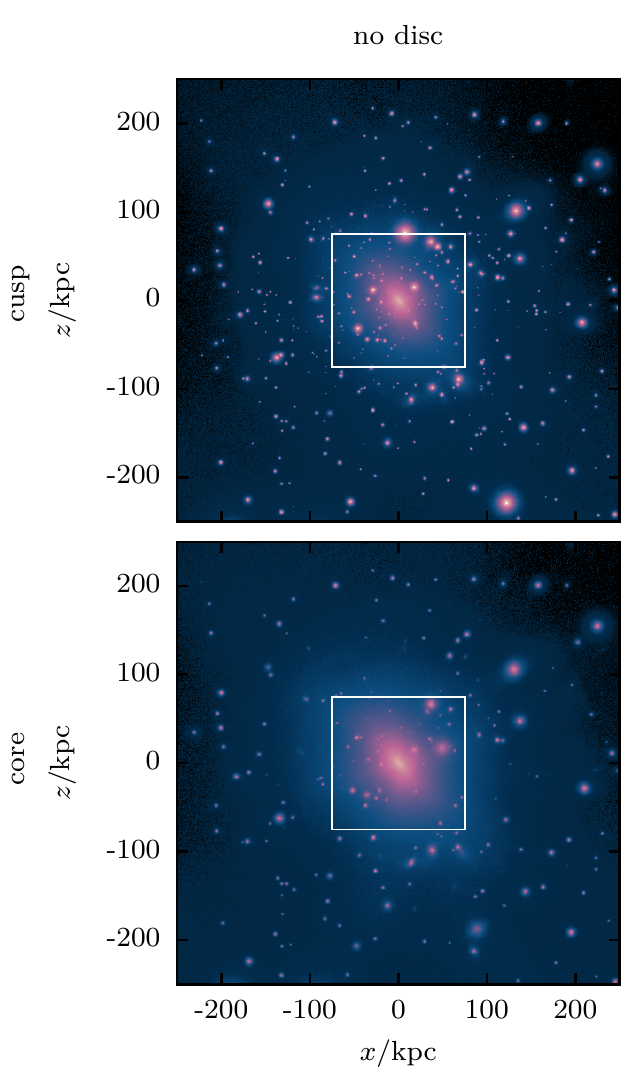} \hfill \includegraphics{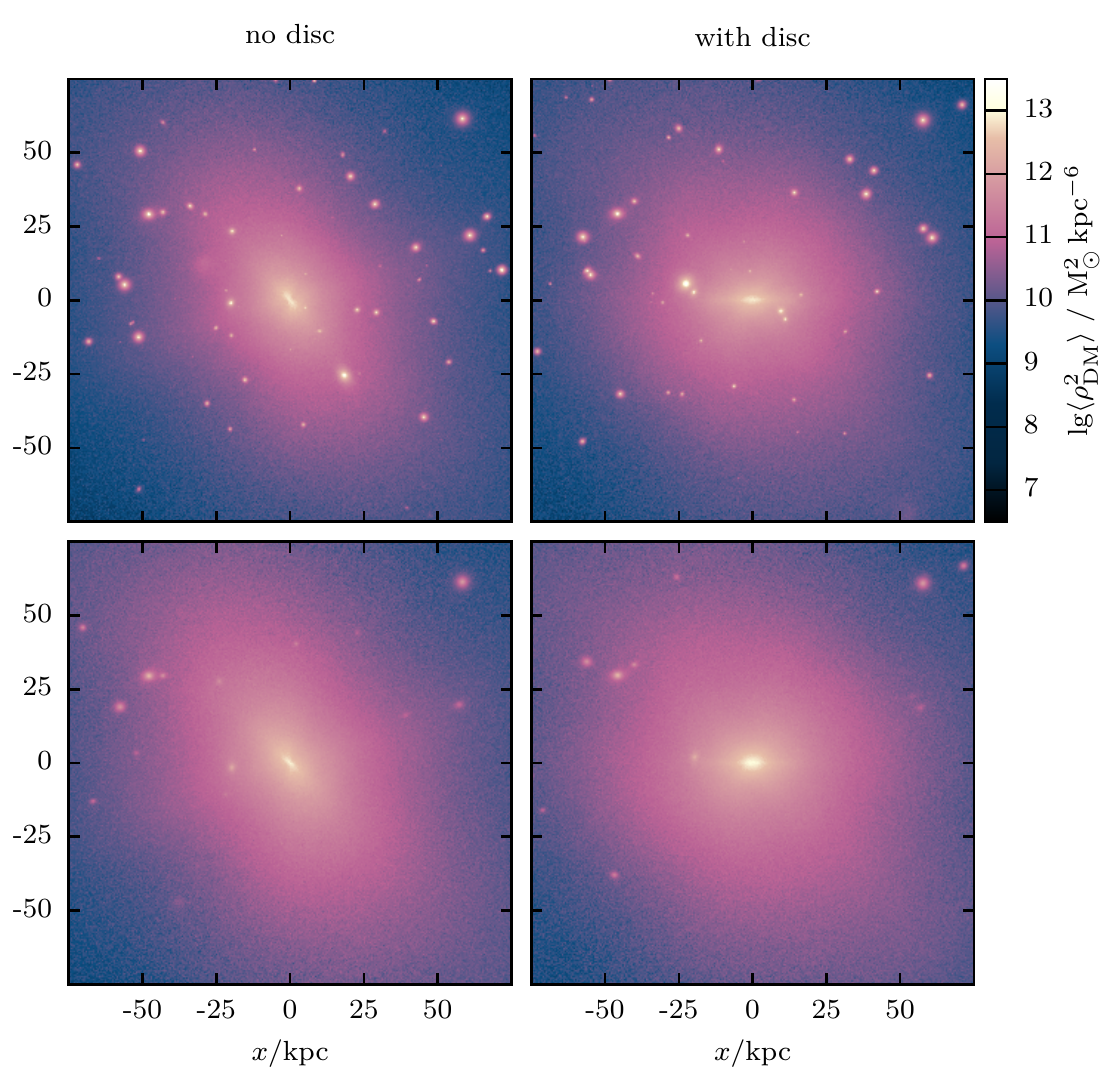}
  \end{minipage}
  \caption{Average squared DM densities at $z=0$ for re-simulations of the Aq-A-2 merger tree in an evolving host halo for cuspy (top row) and cored (bottom row) satellite models. 
  The left (middle) column shows cubic boxes of width $\unit[500]{kpc}$ ($\unit[150]{kpc}$) for a spherical host halo model, while the right column shows edge-on projections of simulations with an axisymmetric host halo, which consists of a spherical NFW halo and a disc. The density is evaluated coarse grained in cubic cells of width $\unit[1]{kpc}$ ($\unit[0.5]{kpc}$) for the $\unit[500]{kpc}$ ($\unit[150]{kpc}$) panels.}
  \label{fig:75kpc}
\end{figure*}

\vspace*{-0.6cm}
\section{Numerical Method}
\label{sec:numMet}
Our controlled simulation is based on the merger history of the Aquarius A-2 run \citep{Springel2008}. The Aq-A-2 run uses a particle mass of $m_p = \unit[1.37\times10^{4}]{M_\odot}$ and $\sim 532\times 10^6$ high-resolution particles in total, with a gravitational softening length of $\epsilon = \unit[65.8]{pc}$, and leads to the formation of a MW-like main halo of virial radius $r_{200}=\unit[246]{kpc}$ and mass  $M_{200}=\unit[1.84\times10^{12}]{M_{\odot}}$.

\noindent\textbf{Host galaxy:}
We model the DM halo of the host galaxy through a \citet{nfw1997} profile. 
The time evolution of the NFW halo follows the analytically motivated functions of \citet{buisthelmi14}, fitted to the Aq-A-2 main halo.
Assuming a Maxwellian distribution of velocities, we approximate the effect of dynamical friction through the \citet{Chandrasekhar43} model.
We compare the orbital decay induced by our semi-analytical model to the orbits of selected satellites in the Aq-A-2 simulation, and find that a Coulomb-logarithm of $\ln \Lambda = 2.2$ reproduces the Aquarius results well, consistent with the findings of \citet{penarrubia02}.
Models including a disc consist of an additional axisymmetric \citet{miyamoto1975} profile,
with radial and vertical scale lengths $a_d = \unit[3.5]{kpc}$ and $b_d = \unit[0.3]{kpc}$ and mass $M_d = 0.1~M_{200}(z)$, where we denote by $M_{200}(z)$ the virial mass of the NFW halo at redshift $z$.
To model dynamical friction in the presence of a disc, for simplicity we require at every radius an isotropic distribution of velocities in a reference frame corotating with the disc, with dispersion and rotation velocity as in \citet[equations 34 and 38]{miyamoto1976} and use the same Maxwellian model of dynamical friction as for the halo.
We keep the joint mass of the disc and halo equal to the Aq-A-2 halo mass at each redshift. This choice of adding a disc potential does not alter the distribution of galactocentric distances of the satellite's first pericentric passages (see green curves in Fig.~\ref{fig:cumuperi}). Subsequent pericentric distances are smaller in the models including a disc: orbits decay faster due to the increased dynamical friction inside the disc, and the re-distribution of orbital energy to tidally stripped material becomes more important with the increased mass loss in the models including a disc.

\noindent\textbf{Accreted satellites: }
we extract positions, velocities and masses from the Aq-A-2 merger tree for all satellites with $M_{200} \geq \unit[10^8]{M_\odot}$ (960 in total). About and below this mass scale, the heating of the intergalactic medium during H\textsc{i} re-ionization prevents gas from cooling, and no stars form \citep{Gnedin2000}. Substructures accreted in groups are modelled as individual satellites for simplicity.
 We use the median \citet{prada2012} relation to estimate the scale radii of the accreted satellites as a function of their mass, which range from $\unit[0.6]{kpc}$ at $ \unit[10^8]{M_\odot}$ to $\unit[8.8]{kpc}$ for the most massive satellite with $ \unit[7.3 \times 10^{10}]{M_\odot}$.
In the same spirit as \citet{Bullock2005}, we model each satellite with a fixed number of $2\times10^6$ particles independent of its mass. This allows us to follow the stripping and dynamical evolution of lower mass substructures more reliably than previously done.
We create equilibrium $N$-body realizations drawn from a modified \citet{hernquist1990} profile with DM scale radius $a_s$ and core size $r_c$: 
\begin{equation}
 \rho(x) = \frac{\rho_c}{ \left( x + c \right) \left( x + 1 \right)^{3} },~\mathrm{where}~c = \frac{r_c}{a_s} ~ , ~ x = \frac{r}{a_s}~.
 \label{equ:herndensity}
\end{equation}
We translate the scale radii of NFW profiles to cuspy Hernquist profiles ($c=0$) noting that ${\partial \ln \rho(x)}/{\partial \ln x} = -2$ for $x=1/2$. In this first study, we also choose a fixed relative core size of $r_c = 0.8\,a_s$, where the order of magnitude of this value is motivated by the estimates of DM scale radius $a_s\sim \unit[1.4]{kpc}$ \citep{penarrubia2008local} and core size $r_c=\unit[1^{+0.8}_{-0.4}]{kpc}$ \citep{AmoriscoAgnelloEvans2013} for the Fornax dSph galaxy. Different DM particle and feedback models suggest a range of values of relative or absolute core size - for simplicity, we choose a constant ratio $r_c/a_s$ for the entire satellite population (see discussion).
The $N$-body models are injected into the host potential at $z_\mathrm{infall}$, defined by the peak of their mass evolution, i.e. before they lose mass due to tidal stripping.
The tidal interaction of the satellites with the host is simulated individually, i.e. encounters between satellites are neglected. 

\noindent\textbf{PM code:}
we follow the evolution of the $N$-body models in the tidal field of the host galaxy with the particle mesh code \textsc{superbox} \citep[see][]{Fellhauer2000}, which samples the density of each satellite to three cubic grids with different resolution and performs a leapfrog integration to solve the equation of motion. The three grids help to account for the large dynamic range in density between the satellites and the tidally stripped material. Each grid consists of $64^3$ cubic cells. Grids 1 and 2 move with the centre of density of each satellite, while grid 3 is centred on the host galaxy. Grid 1 (2, 3) resolves the core of the satellites with a resolution of $2a_s/64$ ($20a_s/64$, $\unit[1]{Mpc}/64$). As the self-gravity of the stripped material is not significant for its evolution, the relatively low resolution of the outermost grid does not have any significant impact on our analysis. We use a time step of $\Delta t = \min( t_c(a_s)/400,~\unit[1]{Myr})$, where $t_c$ is the period of a circular orbit at $a_s$. All models are run in isolation for \unit[14]{Gyrs} in order to test the stability of our numerical set-up, and the profiles are stable within the resolution limits of grid 1. We adopt the same cosmological parameters as the Aquarius project, i.e. $\Omega_m  = 0.25,~ \Omega_\Lambda = 0.75$ and $h = 0.73$.

\vspace*{-0.6cm}
\section{Controlled simulations}
\label{sec:ControlledSims}
In this first contribution, we study the effect of an axisymmetric disc on the abundance of DM substructures at $z=0$. We model the accretion on to the host halo separately for the case of satellites with cuspy and cored DM profiles. Fig.~\ref{fig:75kpc} shows the average squared DM density at $z=0$ for a host galaxy consisting of a spherical NFW halo compared to a host galaxy with an added axisymmetric disc.
Results obtained with a cuspy and a cored satellite population are shown in the top and bottom panels, respectively.
The number of surviving substructures decreases for the models in the order \emph{`cusp - no disc'}, \emph{`cusp - disc'}, \emph{`core - no disc'}, \emph{`core - disc'}, and the difference between the models is more pronounced in the central regions of the galaxy. 
\begin{figure}
  \centering
  \includegraphics[width=8.5cm]{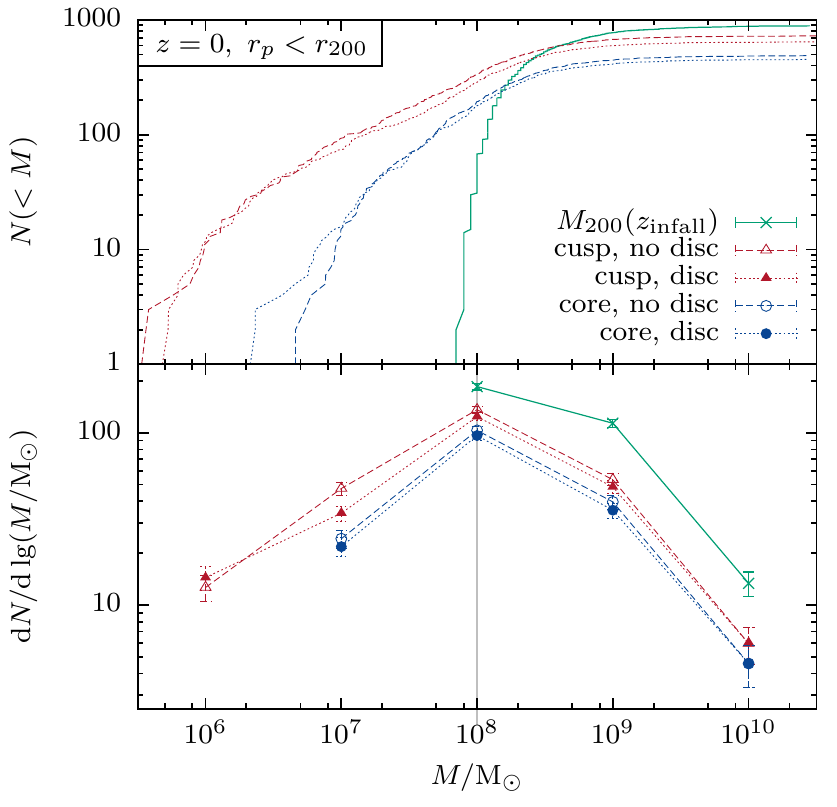} 
  \caption{\emph{Top panel:} number $N(<M)$ of surviving substructures with mass $<M$ at $z=0$. The masses have been obtained by fitting iteratively Hernquist profiles to the particles of each individual satellite. The green solid line shows $N(<M_{200})$ at $z_\mathrm{infall}$. \emph{Bottom panel:} number $\mathrm{d}N/\mathrm{d}\lg(M/\unit{M_\odot})$ of substructures in five logarithmic mass bins, with Poisson errorbars. We do not show bins containing less than 1 per cent of the satellite galaxy population. Note the cut-off in the initial conditions for $M_{200} < \unit[10^8]{M_\odot}$.}
  \label{fig:masshist}
\end{figure}

To characterize the properties of the surviving satellites, we fit spherical Hernquist profiles to the particles of each individual satellite, about the centre of density as determined by \textsc{superbox}. We perform iterative fits, removing particles lying further out than twice the fitted scale radius to reduce the impact of tidally stripped material on the fit parameters, and leave the outer slope of the profile as a free parameter to account for a steeper drop in density induced by tidal stripping \citep[see][]{Penarrubia2010}. Only those fitted profiles which have a central density larger than the background density are considered to be surviving substructures, and we only consider satellites which have a first pericentric radius $r_p<r_{200}(z=0)$.
We find that the satellite DM profiles are altered by tides predominantly in the outskirts, consistent with the findings of \citet{Penarrubia2010}. The evolution of the profile parameters follows \emph{tidal tracks} (i.e. mono-parametric functions which only depend on the fraction of mass lost to tides) close to those of \citet{Penarrubia2010}.

The top panel of Fig.~\ref{fig:masshist} shows that the number $N(<M)$ of surviving substructures at $z=0$ for each given mass $M$ is 
larger for the cuspy models compared to the cored models, and larger for the models without disc compared to those with an added
disc. Note that we do not identify any surviving cored substructures at $z=0$ with $M \lesssim  \unit[2.5 \times 10^6]{M_\odot}$. 
Substructures with $M \lesssim \unit[10^8]{M_\odot}$ at $z=0$ predominantly originate from accreted satellites with masses of $\unit[10^8 \sim 10^9]{M_\odot}$ at $z_\mathrm{infall}$ on orbits with $r_p \lesssim \unit[50]{kpc}$.
Cored satellites, having a shallower gravitational potential at fixed $M_{200}$, are less resilient to tidal stripping  and lose more mass for a fixed number of pericentric passages than their cuspy counterparts \citep[see][Fig.~3]{Penarrubia2010}. 
For the models with added disc, the strong gradient in density perpendicular to the disc plane causes increased tidal forces on the satellites, which in turn get stripped more efficiently. 
In the bottom panel of Fig.~\ref{fig:masshist}, we show that for $M \gtrsim \unit[10^7]{M_\odot}$, the shape of the mass spectrum  $\mathrm{d}N/\mathrm{d}\lg(M/\unit{M_\odot})$ of the surviving substructures does not change between the models, but the normalization decreases, consistently with the visual impression of Fig.~\ref{fig:75kpc}. Models with cored satellites have $\sim 2$ times less surviving substructures than their cuspy counterparts, and models including a galactic disc have $\sim 20$ per cent less substructures compared to models without a disc.\looseness=-1

\begin{figure}
  \centering
  \includegraphics[width=8.5cm]{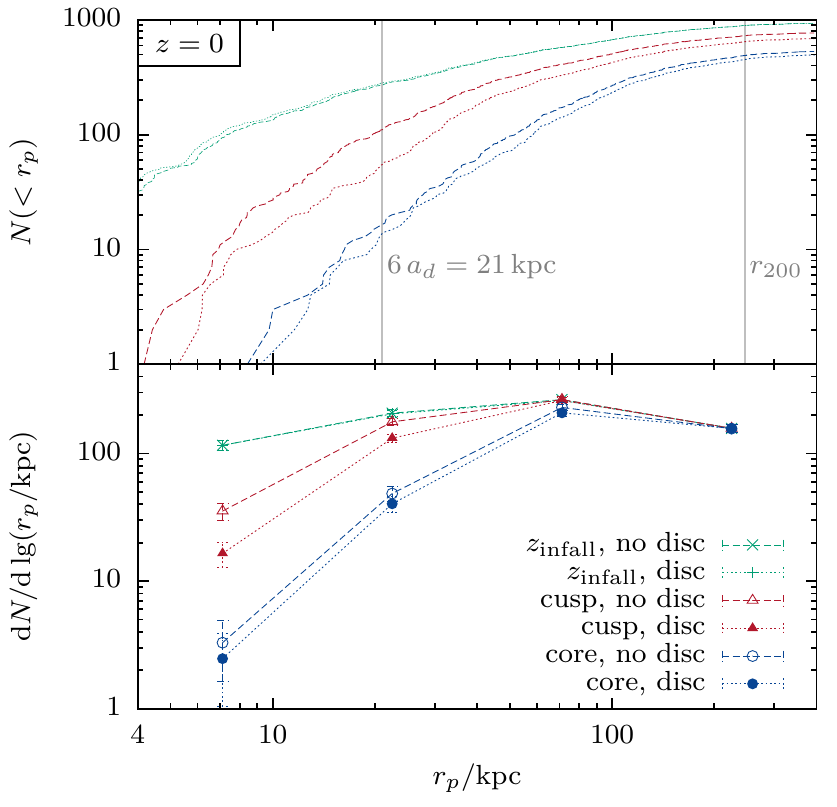} 
  \caption{\emph{Top panel:} number $N(<r_p)$ of surviving substructures with galactocentric distance of the first pericentre $< r_p$ at $z=0$. The green curves show $N(<r_p)$ at $z_\mathrm{infall}$ for models with and without a disc, identical for cuspy and cored models. A radius of $\unit[21]{kpc}$, equal to six times the horizontal disc scale length~$a_d=\unit[3.5]{kpc}$, and the virial radius $r_{200}=\unit[246]{kpc}$ are denoted by solid vertical lines. \emph{Bottom panel:} number $\mathrm{d}N/\mathrm{d}\lg(r_p/\unit{kpc})$ of substructures in four logarithmic pericentre bins, with Poisson errorbars.}
  \label{fig:cumuperi}
\end{figure}

To further examine the process driving the depletion of substructures, we study the abundance of substructures as a function of the galactocentric distance $r_p$ of their first pericentric passage.
The top panel of Fig.~\ref{fig:cumuperi} shows that the number $N(<r_p)$ of surviving substructures at each given $r_p$ is larger for the cuspy models compared to the cored models, and larger for the models without disc compared to those with an added disc. There are no cored substructures with $r_p \lesssim \unit[8]{kpc}$. 
As the tidal forces increase towards the galactic centre, satellites with pericentric distances within a couple of disc scale radii $a_d$ disrupt more easily. Adding to this, satellites with small $r_p$ are generally accreted at earlier redshifts and went through more pericentre passages along their orbits. This explains the reduced number of substructures at small $r_p$, while satellites with large $r_p$ went through fewer pericentric passages during which they experienced weaker tidal forces. 
The bottom panel of Fig.~\ref{fig:cumuperi} shows the number $\mathrm{d}N/\mathrm{d}\lg(r_p/\unit{kpc})$ of surviving substructures in four logarithmic bins of pericentric distance $r_p$. For large $r_p$, where tidal forces and the effect of the disc become negligible, the number of satellites per pericentre bin converges to similar values for all four models. In contrast the difference between the models is largest for satellites on orbits which penetrate the disc, i.e. $r_p \lesssim 6\,a_d$ (factor $\sim 4$ between cuspy and cored models, factor $\lesssim 2$ between models with and without disc).

\vspace*{-0.6cm}
\section{Discussion and conclusions}
\label{sec:Conclusions}
In this Letter, we studied the effect of a galactic disc on the abundance of cuspy and cored DM substructures in an evolving MW-like halo. 
We re-simulated the Aquarius Aq-A-2 merger tree by following the evolution of individual satellites, modelled as equilibrium $N$-body realizations of generalized Hernquist profiles with a fixed number of $2\times 10^6$ particles and injected in the host potential at the peak of their mass evolution.
We show that models with cuspy satellites have twice as many surviving substructures at $z=0$ than their cored counterparts for pericentres $r_p < r_{200}$, and four times as many for $r_p < 0.1\,r_{200}$. Adding an evolving axisymmetric disc of mass $0.1\,M_{200}(z)$ and radial scale length $a_d=\unit[3.5]{kpc}$ to the potential reduces the number of both cuspy and cored substructures further by a factor $\lesssim 2$ for satellites on orbits which penetrate the disc, and by $\sim 20$ per cent for the entire population.
Remarkably, we find a factor $\sim 8$ depletion in the number of cored substructures which penetrate the disc with respect to their cuspy counterparts evolving in a DM-only halo without a disc.
Our results on the depletion of substructures due to the disc are in good agreement with the findings of previous studies: \citet[][cf. Fig.~3]{DOnghia10} show a depletion by a factor of $\sim 2$ for for satellites with $M\sim\unit[10^9]{M_\odot}$ located at $z=0$ within the inner $\unit[30]{kpc}$ of a MW-like halo, whereas \citet[][cf. Fig.~18]{YurinSpringel2015} and \citet[][cf. Fig.~4]{ZhuMarinacci2016} find a depletion of the order of $\sim 30$ per cent for the entire population.\looseness=-1

In contrast to self-consistent cosmological simulations with fixed $N$-body particle mass and gravitational softening length, our re-simulation technique follows the dynamical evolution of accreted substructures with the same numerical resolution for systems spanning many orders of magnitude in mass and size.
For comparison, the Aq-A-2 run has a fixed particle mass of $m_p = \unit[1.37 \times 10^4]{M_\odot}$, consequently haloes with $M_{200} = \unit[10^8]{M_\odot}$ consist of $\sim 10^4$ particles at infall. Tidal stripping then reduces the particle number further. The simulations adopted in the studies of \citet{DOnghia10}, \citet{YurinSpringel2015} and \citet{ZhuMarinacci2016} use even larger DM particle masses of $m_p > \unit[10^5]{M_\odot}$.
For haloes with a mass of $\unit[10^8]{M_\odot}$ at $z_\mathrm{infall}$, our satellites contain $\sim 2$ orders of magnitude more particles than the Aq-A-2 subhaloes.
Furthermore, modelling the cosmological evolution of the host-halo and disc potential analytically achieves high-enough resolution at a modest computational cost, which allows us to perform controlled simulations where we change single parameters at a time and quantify their effect on the surviving satellite population.

Although the re-simulation technique is computationally highly efficient, several aspects of the set-up call for caution. 
(i)~Simulating the tidal interaction of individual satellites with the host halo does not account for the interaction between satellites observed in group infall. Also, satellites contain themselves sub-substructures, which we also model as individual satellites. 
(ii)~For simplicity, our current host DM halo is approximated by an evolving spherical potential, thus neglecting the impact of triaxiality.
However, \citet{GillKnebe2004} find that the destruction rate of substructures is nearly independent of the triaxiality of the host halo.
(iii)~The disc is modelled with an isotropic velocity dispersion in the corotating frame, which leads to lower rotation velocities compared to the MW disc. Consequently the difference in efficiency of dynamical friction on satellites with pro- and retrograde orbits is attenuated.
(iv)~Regarding the internal DM distribution of the satellites, in this first contribution we consider a fixed ratio of DM scale radius $a_s$ and core size $r_c = 0.8\,a_s$, while different DM particle and feedback models motivate many other choices for relative or absolute core sizes \citep{Pontzen2012, Vogelsberger2012, Rocha2013, DiCintio2014, Elbert2015}. In a future contribution, we explore these choices by constructing satellite libraries with a range of different core sizes.

This Letter introduces the \emph{Cusp Core Comparison Project}, a suite of re-simulations of MW-like haloes specifically tailored to study the abundance of satellites and the properties of DM and stellar haloes resulting from changes to the internal DM density profiles of accreted substructure. Among several goals, we aim to study the dynamical properties of DM (e.g. DM caustics - see \citealt{VogelsbergerWhite2011}, the annihilation signal expected from dark subhaloes - see e.g. \citealt{DiemandKuhlenMadau2007}, \citealt{SpringelWhiteNature2008}, \citealt{Vogelsberger2009} ), the structure of stellar haloes (e.g. abundance and spatial distribution of ultra-faints, and number of streams in the solar neighbourhood - see \citealt{FacundoGomez2013}, \citeyear{FacundoGomez2014}), and how these relate to observations in the MW.
By convolving our models with the expected \emph{Gaia} errors and completeness, our models may provide a flexible tool to test a range of different DM particle models and to study the hierarchical assembly of galaxies using the MW as a test bed.

\vspace*{-0.6cm}
\section*{Acknowledgements}
\urlstyle{rm}
This work used the ARCHER UK National Supercomputing Service (\url{http://www.archer.ac.uk}), and the authors would like to thank the administrators for their support. We thank the VIRGO Consortium for giving us access to the Aquarius trees.
The authors would like to thank Robyn Sanderson and the anonymous referee for their constructive comments.
CFPL is supported by a Junior Fellow award from the Simons Foundation.
RE acknowledges support through the Scottish University Physics Alliance.

\goodbreak
\renewcommand{\baselinestretch}{0.99}
\footnotesize{
\bibliography{cccp1}
}
\vspace*{-0.4cm}
\bsp	
\label{lastpage}
\end{document}